# BRAIN-ADAPTER: ENHANCING NEUROLOGICAL DISORDER ANALYSIS WITH ADAPTER-TUNING MULTIMODAL LARGE LANGUAGE MODELS

*Jing Zhang[1], Xiaowei Yu[1], Yanjun Lyu[1], Lu Zhang[2], Tong Chen[1], Chao Cao[1], Yan Zhuang[1], Minheng Chen[1], Tianming Liu[3], Dajiang Zhu[1]*

[1]Computer Science and Engineering, The University of Texas at Arlington, Arlington, TX, USA
[2]Department of Computer Science, Indiana University Indianapolis, IN, USA
[3]School of Computing, The University of Georgia, Athens, GA, USA

## ABSTRACT

Understanding brain disorders is crucial for accurate clinical diagnosis and treatment. Recent advances in Multimodal Large Language Models (MLLMs) offer a promising approach to interpreting medical images with the support of text descriptions. However, previous research has primarily focused on 2D medical images, leaving richer spatial information of 3D images under-explored, and single-modality-based methods are limited by overlooking the critical clinical information contained in other modalities. To address this issue, this paper proposes Brain-Adapter, a novel approach that incorporates an extra bottleneck layer to learn new knowledge and instill it into the original pre-trained knowledge. The major idea is to incorporate a lightweight bottleneck layer to train fewer parameters while capturing essential information and utilize a Contrastive Language-Image Pre-training (CLIP) strategy to align multimodal data within a unified representation space. Extensive experiments demonstrated the effectiveness of our approach in integrating multimodal data to significantly improve the diagnosis accuracy without high computational costs, highlighting the potential to enhance real-world diagnostic workflows.

## 1. INTRODUCTION

Neurological disorders like Alzheimer's Disease (AD), profoundly disrupt individuals' social, linguistic, and cognitive abilities, posing significant public health challenges globally [1]. While there is currently no definitive cure for AD once established, early diagnosis is critical for enabling timely intervention and delaying the disease's progression. Over the last decade, with the advancements in machine learning (ML) technologies, researchers have applied various ML methods like Graph Convolutional Neural Networks (GCN) [2, 3] and transformers [4, 5] to distinguish individuals with normal cognition (NC), mild cognitive impairment (MCI), and AD. Despite significant progress, previous studies have predominantly focused on neuroimaging data, such as Magnetic Resonance Imaging (MRI) scans and Position Emission Tomography (PET), overlooking the inherently intertwined nature of images and text in clinical diagnosis and treatment.

On the one hand, the recently revised clinical criteria for detecting AD emphasize the importance of using multiple modalities, such as core biomarkers and clinical tests, for diagnosis [6]. With the widespread implementation of electrical health records (EHRs), a wealth of routine clinical reports, including those provided by the Alzheimer's Disease Neuroimaging Initiative (ADNI), are digitally available. These reports offer rich details including demographic attributes, biomarker measurements, and cognitive and neurofunctional assessments, making them an invaluable resource for incorporating into brain disease studies.

On the other hand, MLLMs have recently demonstrated remarkable performance in various multimodal tasks by effectively integrating and mapping images and text into a joint space [7, 8, 9, 10, 11]. Inspired by the success of MLLMs, researchers believe that knowledge jointly learned from medical images and reports could be beneficial for downstream clinical tasks such as classification [12, 13], segmentation [14, 15, 16], medical visual question answering (VQA) [17, 18], and report generation [19, 20], leading to considerable research efforts in this area. However, adopting MLLMs for diagnosing neurological disorders presents several significant challenges:

- The rich spatial information in 3D medical images like MRIs, while valuable, poses significant computational and analytical challenges compared to two-dimensional natural images.
- Most existing MLLMs typically process text as corresponding descriptions of paired images. However, in diagnosing AD, physicians treat images and reports as complementary, as reports provide critical information beyond what is visible in medical scans.
- The inherent high dimensions and numerous data points of 3D images, coupled with the relatively limited scale of medical data, pose another challenge. Although collecting large-scale, high-quality datasets and training task-specific models provide a feasible solution, it incurs significant human and computational costs.

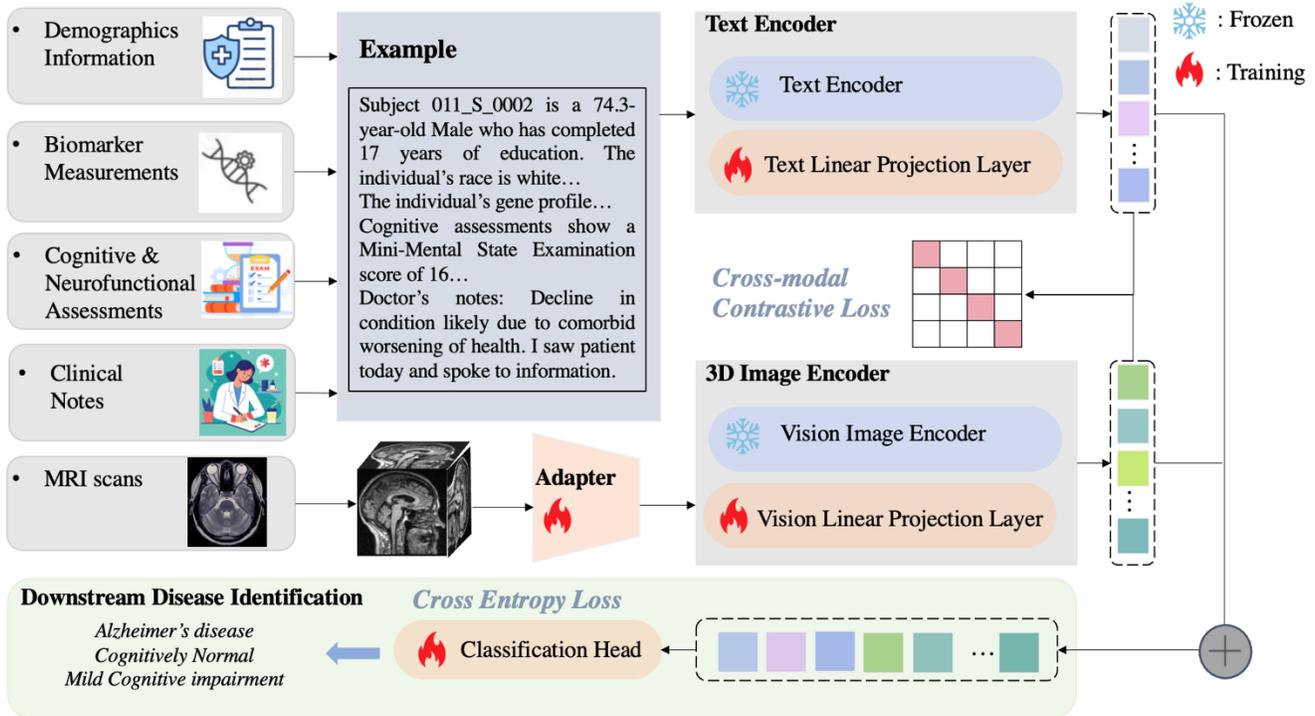

**Fig.1.** Overview of the proposed adapter-tuning MLLM framework: the image and text encoder are frozen while the trainable Adapter and Linear Projection Layer are updated. It integrates the fine-tuned newly acquired brain disease-related knowledge with the original medical knowledge inside the MLLM for the downstream tasks: brain disease identification.

To address the above challenges, in this work, we propose Brain-Adapter as shown in Fig. 1, a lightweight bottleneck architecture that fine-tunes only a small number of additional parameters. First, inspired by the adapter modules in parameter-efficient transfer learning [21, 22], we designed a learnable embedding adapter to bridge the gap between brain MRI scans and a pre-trained image encoder, which significantly reduces the amount of image data required for learning. Next, we conducted the CLIP training strategy [23], to align different types of medical data within a common space and generate robust representations for downstream prediction tasks. Lastly, we adopted M3D [24], a medical-domain pre-trained MLLM as the backbone of our proposed model. To achieve efficient fine-tuning, the parameters of the core encoders are frozen except linear probe during the fine-tuning process. Our model is lightweight and trainable on a single NVIDIA A6000 GPU, which can be easily applied to current clinical settings. In summary, our contributions are as follows:

- Establish comprehensive image-text pairs from ADNI, which include demographic information, biomarker measurements, cognitive assessments, and unstructured clinical notes. Furthermore, we apply multiple preprocessing methods to the images, enhancing the robustness of our framework.
- Propose Brain-Adapter, which leverages both the knowledge embedded in the original MLLMs and the newly acquired knowledge from new training examples, while minimizing the number of parameters and reducing computational costs.
- By aligning the understanding of brain image with corresponding clinical reports, our framework effectively leverages these complementary sources of information, demonstrating promising performance in distinguishing AD and MCI subjects from NC.

## 2. METHOD

### 2.1. Data Collection

Initially, we collected tabular EHRs from ADNI and analyzed the key variables. These variables were then converted into natural language clinical reports using a key-value pair template to align with the sequential nature of the language model. Subsequently, we gathered corresponding 16797 post-processed T1-weighted MRI scans from patients in the ADNI. Finally, we constructed a multimodal dataset specifically for Alzheimer's disease research.

From the text perspective, the clinical details include demographic information (e.g., age, sex, educational level, etc); biomarker measurements (e.g., allele producing the apolipoprotein APOE-4); cognitive assessments (e.g., Mini-Mental State Examination, Clinical Dementia Rating Scale, etc); as well as doctor's notes (e.g., "Decline in condition likely due to comorbid worsening of health."), which are unstructured.

From the image perspective, the T1-weighted MRI scans provide crucial features. For example, post-mortem investigations have shown that structures in the medial temporal lobe, particularly the entorhinal cortex and hippocampus, are the first to exhibit changes in AD and can be observed through MRI imaging techniques. Additionally, the ADNI dataset provides multiple post-processed versions of these scans using different processing methods such as "MPR; GradWarp; B1 Correction; N3; Scaled," generated from different labs. To enhance the Robustness of our model, we utilize all these processing methods rather than relying on one single specific method. To increase the homogeneity of the data, we resampled the images to a resolution of $256 \times 256 \times 256$ and normalized the intensity to the range $[0,1]$.

In this work, the dataset includes 4,661 AD scans, 5,025 CN scans, and 7,111 MCI scans, with 70% used for training and 30% for testing.

## 2.2 Overall Framework

In our study, we adopt the M3D [26], a model designed for 3D medical image analysis following a CLIP-like strategy, as the backbone of our model to avoid training from scratch. To achieve our goal, we need to address two key challenges: (1) bridging the domain and dimension gaps between the pre-trained vision encoder and our medical scan, and (2) aligning image features with textural ones, allowing all inputs mapped into the latent space for learning multi-modal representations. Inspired by the CLIP-Adapter [24], we propose a Brain-Adapter based on a CNN architecture as Fig. 2, which maps the raw 3D images into a lower-dimensionality vector and aligns with the input token size for the M3D image encoder. Specifically, given an input volume $x_u$ with dimensions $X_u \in \mathbb{R}^{256 \times 256 \times 256}$, the Brain-Adapter consists of a series of convolutional layers with a residual block, resulting in a reduced dimension 3D image $Z_x \in \mathbb{R}^{32 \times 256 \times 256}$:

$$Z_x = \phi_{Adapter}(x) \qquad (1)$$

The M3D's vision encoder adopts a 3D ViT to segment the T1 images into non-overlapping patches of size $P = 4 \times 16 \times 16$, resulting in patches $\{P_i\}_{i=1}^{N}$. Each patch is then mapped into a $256 \times 768$ dimensional space D, representing 256 tokens with 768 feature dimensions. To minimize the computational cost, we kept the vision encoder's parameters frozen, updating only the linear projection layer during fine-tuning. The M3D text encoder utilizes LLaMA2-7B, a model that has demonstrated effectiveness in capturing linguistic patterns across various medical domains. The core text encoder module was kept frozen, while only the linear projection layer was updated.

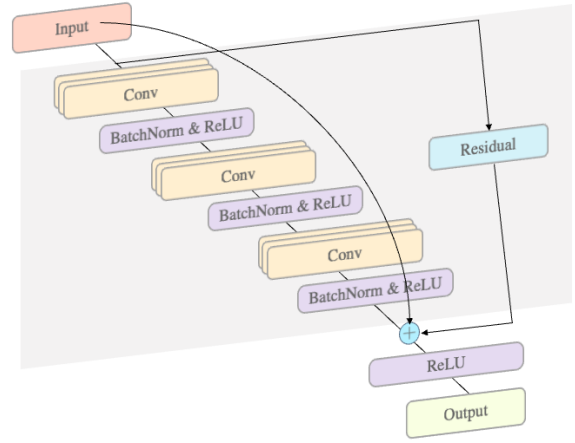

**Fig.2.** Illustration of proposed Brain-Adapter, a lightweight residual-style convolutional structure adapter before the MLLM's vision encoder.

## 2.3 Loss Function

To achieve cross-modality interactions, one of our framework fine-tuning objectives is cross-model contrastive loss. During the fine-tuning stage, given a batch of $N$ input pairs $(x_u, x_v)$ from training data, we first calculate their corresponding representation pairs $(u, v)$. Let $(u_i, v_i)$ denote the $i$-th pair, then, the image-to-text contrastive loss for the $i$-th pair is defined as:

$$\mathcal{L}_{u \to v}^i = -\log \frac{exp(\langle u_i, v_i \rangle / \tau)}{\sum_{k=1}^{N} exp(\langle u_i, v_k \rangle / \tau)} \qquad (2)$$

Similarly, the text-to-image contrastive loss for the i-th pair is:

$$\mathcal{L}_{v \to u}^i = -\log \frac{exp(\langle v_i, u_i \rangle / \tau)}{\sum_{k=1}^{N} exp(\langle v_i, u_k \rangle / \tau)} \qquad (3)$$

Here, $\tau \in \mathbb{R}^+$ is a learnable temperature parameter to scale logits, and $\langle u, v \rangle$ represents the cosine similarity:

$$\langle u, v \rangle = \frac{u^T v}{\|u\| \|v\|} \qquad (4)$$

One of our goals is to align the brain image and the corresponding clinical reports within a common representational space. To achieve this, our method computes a cross-modal contrastive loss. This final loss is computed as a combination of (2) and (3):

$$\mathcal{L}_{contrastive} = \frac{1}{2N} \sum_{i=1}^{N} (\mathcal{L}_{u \to v}^i + \mathcal{L}_{v \to u}^i) \qquad (5)$$

Next, to achieve the downstream task of brain disease prediction, the image and text representations generated from the MLLM were combined and passed through a single-layer perceptron with a nonlinear activation function, which serves as the prediction head. The objective is to guide the model to improve its performance by minimizing the cross-entropy loss. Finally, we optimize a joint loss function $\mathcal{L}_{cls}$ that combines $\mathcal{L}_{contrastive}$ and $\mathcal{L}_{CE}$, balanced by learnable hyperparameters $\lambda_1$ and $\lambda_2$:

$$\mathcal{L}_{cls} = \lambda_1 \mathcal{L}_{contrastive} + \lambda_2 \mathcal{L}_{CE} \qquad (6)$$

## 3. EXPERIMENT

### 3.1. Experimental Setting

The fine-tuning process was conducted over 9 epochs with a batch size of 8, utilizing a single NVIDIA A6000 GPU. We employed the AdamW optimizer, setting the learning rate for the Brain-Adapter to $1e-3$ and for the pre-trained MLLM to $1e-4$.

### 3.2. Classification Results

We selected 3D ResNet50 [25] and 3D DenseNet121 [26] as baselines. Since these are single-modal image encoders, for a fair comparison, we used only the features extracted by the image encoder for the brain disease classification task. Unlike previous works that primarily focus on NC/MCI identification [2, 3], our model addresses a three-class classification task, and we evaluated the performance of each class individually. As shown in Table 1, by simply unfreezing the linear projection layer, our model achieves the best performance, demonstrating that the inherent knowledge in MLLMs indeed aids in diagnosing neurological disorders with fewer training parameters.

**Table 1.** Comparison of our proposed method and baseline

| Methods | Group | Modality | PRE | SEN | F1 |
|---|---|---|---|---|---|
| 3D ResNet50 | AD | MRI image | 0.62 | 0.71 | 0.66 |
|  | CN |  | 0.58 | 0.72 | 0.64 |
|  | MCI |  | 0.59 | 0.63 | 0.61 |
| 3D DenseNet 122 | AD | MRI image | 0.71 | 0.76 | 0.73 |
|  | CN |  | 0.73 | 0.70 | 0.71 |
|  | MCI |  | 0.69 | 0.77 | 0.72 |
| Ours (FPM) | AD | MRI image | 0.55 | 0.35 | 0.43 |
|  | CN |  | 0.40 | 0.36 | 0.38 |
|  | MCI |  | 0.48 | 0.67 | 0.56 |
| Ours (TLP) | AD | MRI image | 0.73 | 0.76 | 0.75 |
|  | CN |  | **0.78** | 0.69 | 0.73 |
|  | MCI |  | 0.69 | **0.87** | **0.77** |

FPM: Frozen Pre-trained MLLM    TLP: Training Linear projection

Table 2 reports the classification results on our multimodal dataset. When all the parameters of the MLLM are frozen and only the Brain-Adapter is trained, the model can still classify diseases, indicating that the Brain-Adapter effectively leverages the knowledge embedded in the original MLLMs. Moreover, when we simply unfreeze the linear projection layer of both the image encoder and the text encoder, the classification performance improves significantly, showing that the Brain-Adapter can inject new knowledge from fresh training examples into the original MLLMs. Furthermore, the latter shows a significant improvement when we compare Table 1, which uses only a single MRI image modality, with Table 2, which uses multimodal data. This suggests that clinical reports help the model better understand brain diseases and our method aligns brain images with the corresponding clinical reports for a more comprehensive understanding successfully. Fig. 3 shows how the multimodal feature distribution in the latent space changes over training. Initially, CN and MCI representations are close due to their similar clinical features. By the 9th epoch, all three classes become distinct in the latent space with increased separation, demonstrating our framework's effectiveness in brain disease diagnosis.

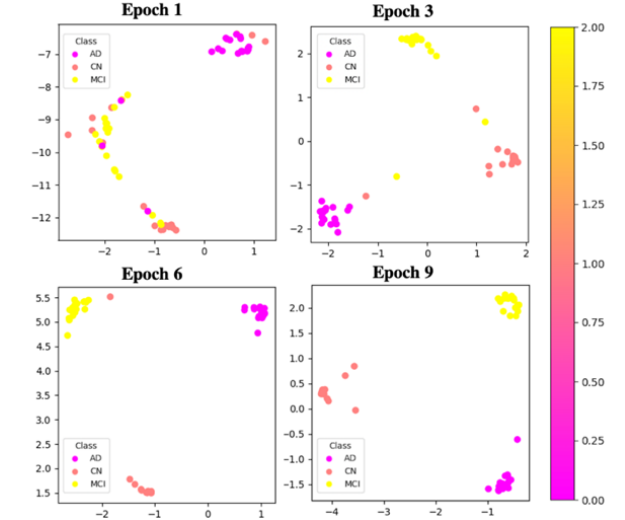

**Fig. 3.** t-SNE visualization of the embeddings

**Table 2.** Ablation of the linear projection layer

| Methods | Group | Modality | PRE | SEN | F1 |
|---|---|---|---|---|---|
| Ours (FPM) | AD | Image-Text pair | 0.60 | 0.40 | 0.47 |
|  | CN |  | 0.50 | 0.39 | 0.43 |
|  | MCI |  | 0.53 | 0.72 | 0.61 |
|  | M-Avg |  | 0.54 | 0.50 | 0.51 |
|  | W-Avg |  | 0.54 | 0.53 | 0.52 |
| Ours (TLP) | AD | Image-Text pair | 0.92 | **0.94** | **0.93** |
|  | CN |  | **0.96** | 0.81 | 0.88 |
|  | MCI |  | 0.87 | **0.94** | 0.90 |
|  | M-Avg |  | 0.92 | 0.90 | 0.90 |
|  |  |  | ↑0.38 | ↑0.40 | ↑0.39 |
|  | W-Avg |  | 0.91 | 0.91 | 0.91 |
|  |  |  | ↑0.37 | ↑0.38 | ↑0.39 |

M-Avg: Macro-average (Treats all classes equally)
W-Avg: Weighted-average (Weights classes by size)

## 4. CONCLUSION

We propose Brain-Adapter, a lightweight bottleneck architecture that fine-tunes minimal parameters to learn complex 3D MRI images. By leveraging pre-trained MLLM medical knowledge and cross-model contrastive training, our method aligns brain images with clinical reports in a unified representation space. This approach enhances diagnostic accuracy, reduces analysis time, and integrates multimodal data, providing a practical real-world diagnostic workflow.

## 5. ACKNOWLEDGMENTS

This work was supported by National Institutes of Health (R01AG075582 and RF1NS128534)

## 6. COMPLIANCE WITH ETHICAL STANDARDS

This is a computational simulation study for which no ethical approval was required.

## 7. REFERENCES

[1] Feigin V L, Vos T, Nichols E, et al. The global burden of neurological disorders: translating evidence into policy[J]. The Lancet Neurology, 2020, 19(3): 255-265.

[2] Zhang L, Wang L, Gao J, et al. Deep fusion of brain structure-function in mild cognitive impairment[J]. Medical image analysis, 2021, 72: 102082.

[3] Zhang L, Na S, Liu T, et al. Multimodal deep fusion in hyperbolic space for mild cognitive impairment study[C]//International Conference on Medical Image Computing and Computer-Assisted Intervention. Cham: Springer Nature Switzerland, 2023: 674-684.

[4] Yu X, Zhang L, Cao C, et al. Gyri vs. Sulci: Core-Periphery Organization in Functional Brain Networks[C]//International Conference on Medical Image Computing and Computer-Assisted Intervention. Cham: Springer Nature Switzerland, 2024: 164-174.

[5] Yu X, Zhang L, Zhao L, et al. Disentangling spatial-temporal functional brain networks via twin-transformers[J]. arXiv preprint arXiv:2204.09225, 2022.

[6] Carrillo M C, Masliah E. NIA-AA Revised Clinical Criteria for Alzheimer's Disease[C]//Alzheimer's Association International Conference. ALZ, 2023.

[7] Zhao Z, Liu Y, Wu H, et al. Clip in medical imaging: A comprehensive survey[J]. arXiv preprint arXiv:2312.07353, 2023.

[8] Pei X, Zuo K, Li Y, et al. A review of the application of multi-modal deep learning in medicine: bibliometrics and future directions[J]. International Journal of Computational Intelligence Systems, 2023, 16(1): 44.

[9] Wang Z, Wu Z, Agarwal D, et al. Medclip: Contrastive learning from unpaired medical images and text[J]. arXiv preprint arXiv:2210.10163, 2022.

[10] Lei J, Dai L, Jiang H, et al. Unibrain: Universal brain mri diagnosis with hierarchical knowledge-enhanced pre-training[J]. arXiv preprint arXiv:2309.06828, 2023.

[11] Yu X, Wu Z, Zhang L, et al. Cp-clip: Core-periphery feature alignment clip for zero-shot medical image analysis[C]//International Conference on Medical Image Computing and Computer-Assisted Intervention. Cham: Springer Nature Switzerland, 2024: 88-97.

[12] Xue C, Kowshik S S, Lteif D, et al. AI-based differential diagnosis of dementia etiologies on multimodal data[J]. Nature Medicine, 2024: 1-13.

[13] Peng L, Cai S, Wu Z, et al. MMGPL: Multimodal Medical Data Analysis with Graph Prompt Learning[J]. Medical Image Analysis, 2024: 103225.

[14] Hatamizadeh A, Nath V, Tang Y, et al. Swin unetr: Swin transformers for semantic segmentation of brain tumors in mri images[C]//International MICCAI brainlesion workshop. Cham: Springer International Publishing, 2021: 272-284.

[15] Liu C, Ouyang C, Chen Y, et al. T3d: Towards 3d medical image understanding through vision-language pre-training[J]. arXiv preprint arXiv:2312.01529, 2023.

[16] Tang Y, Yang D, Li W, et al. Self-supervised pre-training of swin transformers for 3d medical image analysis[C]//Proceedings of the IEEE/CVF conference on computer vision and pattern recognition. 2022: 20730-20740.

[17] Wang S, Zhao Z, Ouyang X, et al. Interactive computer-aided diagnosis on medical image using large language models[J]. Communications Engineering, 2024, 3(1): 133.

[18] Chen Q, Hu X, Wang Z, et al. Medblip: Bootstrapping language-image pre-training from 3d medical images and texts[J]. arXiv preprint arXiv:2305.10799, 2023.

[19] Yu F, Endo M, Krishnan R, et al. Evaluating progress in automatic chest x-ray radiology report generation[J]. Patterns, 2023, 4(9).

[20] Ma C, Jiang H, Chen W, et al. Eye-gaze Guided Multi-modal Alignment Framework for Radiology[J]. arXiv preprint arXiv:2403.12416, 2024.

[21] Houlsby N, Giurgiu A, Jastrzebski S, et al. Parameter-efficient transfer learning for NLP[C]//International conference on machine learning. PMLR, 2019: 2790-2799.

[22] Gao P, Geng S, Zhang R, et al. Clip-adapter: Better vision-language models with feature adapters[J]. International Journal of Computer Vision, 2024, 132(2): 581-595.

[23] Radford A, Kim J W, Hallacy C, et al. Learning transferable visual models from natural language supervision[C]//International conference on machine learning. PMLR, 2021: 8748-8763.

[24] Bai F, Du Y, Huang T, et al. M3d: Advancing 3d medical image analysis with multi-modal large language models[J]. arXiv preprint arXiv:2404.00578, 2024.

[25] He K, Zhang X, Ren S, et al. Deep residual learning for image recognition[C]//Proceedings of the IEEE conference on computer vision and pattern recognition. 2016: 770-778.

[26] Ruiz J, Mahmud M, Modasshir M, et al. 3D DenseNet ensemble in 4-way classification of Alzheimer's disease[C]//Brain Informatics: 13th International Conference, BI 2020, Padua, Italy, September 19, 2020, Proceedings 13. Springer International Publishing, 2020: 85-96.